\documentstyle[preprint,aps]{revtex}
\begin{document}

\draft
\title{Study of the levels in $^{17}$C above the $^{16}$C + neutron threshold
}

\author{G. Raimann,$^{(1)}$  A. Ozawa,$^{(2)}$ R.N. Boyd,$^{(1,3)}$
F.R. Chloupek,$^{(1)}$  M. Fujimaki,$^{(2)}$ \\ K. Kimura,$^{(4)}$
T. Kobayashi,$^{(2)}$  J.J. Kolata,$^{(5)}$  S. Kubono,$^{(6)}$
I. Tanihata,$^{(2)}$ \\ Y. Watanabe,$^{(2)}$ K. Yoshida$^{(2)}$
}
\address{$^{(1)}$ Department of Physics, The Ohio State University,
Columbus, OH 43210, USA}
\address{$^{(2)}$ The Institute of Physical and Chemical Research (RIKEN),
Wako-shi, Saitama 351-01, Japan}
\address{$^{(3)}$ Department of Astronomy, The Ohio State University,
Columbus, OH 43210, USA}
\address{$^{(4)}$ Nagasaki Institute of Applied Science, Nagasaki,
Nagasaki 851-01, Japan}
\address{$^{(5)}$ Department of Physics, University of Notre Dame, Notre Dame,
IN 46556, USA}
\address{$^{(6)}$ Institute of Nuclear Study, University of Tokyo,
Tanashi, Tokyo 188, Japan}

\maketitle

\begin{abstract}
The $\beta$-delayed neutron decay of $^{17}$B was studied using a radioactive
ion beam.
The neutron energies, measured via
time-of-flight, give information on states in $^{17}$C above the $^{16}$C
+ neutron threshold. States in $^{17}$C were found at excitation energies
of 2.25(2) MeV, 2.64(2) MeV and 3.82(5) MeV, and possibly at 1.18(1) MeV.
These low lying states are of possible interest for
nuclear physics as well as for astrophysics.
\end{abstract}


\section{Introduction}

The Inhomogeneous Models (IMs) of primordial nucleosynthesis
\cite{App88,Alc87,Mal89,Rau94,Jed94} have recently received considerable
attention, especially with regard to abundances predicted for the elements
$^7$Li, $^9$Be and $^{11}$B \cite{Kaj90,Tho94,Jed94b}. Because of the spatial
variation of both the matter density and the ratio of neutrons to protons,
the abundances calculated within the framework of the IMs differ from those
obtained within the Standard Model \cite{Rau94,Jed94}.
Comparison of the predicted primordial
abundances with the observed ones should clarify the level to which
inhomogeneities existed in the early universe.
For various reasons, however, the abundance of $^7$Li may not be a
good measure \cite{Del93}. The primordial abundances of $^9$Be
\cite{Rya90,Gil91} and $^{11}$B \cite{Dun92} are somewhat less sensitive to
unknown factors, but neither is easy to observe.
However \cite{Jed94}, the IMs also predict a considerably
higher abundance of elements with mass 12 and greater than does the
Standard Model. Thus the heavier nuclides might provide an independent
way of testing the predictions of the IMs.
Good candidates, at least from the astronomer's perspective,
appear to be the elements Na, Mg, Al and Si \cite{Rau94}. In
order to establish accurate predictions of elemental abundances
based on the IMs, however, accurate knowledge
of nuclear reaction rates between the intervening nuclides is essential.

Most of the nucleosynthesis in the IMs, however, occurs through nuclei
that are just beyond the
neutron rich side of stability, and thus involves short-lived nuclides,
mostly with half-lives
of the order of a second. In addition, the reactions that are most crucial
to linking these nuclides tend to be neutron radiative capture reactions,
necessitating an indirect mechanism to determine the cross
sections.

Recent studies \cite{Rau94,Kaj90b} on the pathway of nucleosynthesis
anticipated in
IMs have identified some of the key branch points in that flow. Specifically,
masses 16, 17 and 19 are important for synthesizing nuclides heavier than mass
20.
In this work we investigated levels in $^{17}$C above the neutron threshold,
using $\beta$-delayed neutron decay of $^{17}$B. These levels
would play a crucial role for the $^{16}$C(n,$\gamma$)$^{17}$C reaction
that would lead on to yet higher masses, although this reaction is not
precisely
in the main path of the flow.

The nucleus $^{17}$B is of interest not only because of its
$\beta$-delayed multi-neutron emission \cite{Ree91,Duf88}, but also for
its anomalously large matter radius \cite{Tan88,Oza94}. The properties of this
neutron halo have been investigated theoretically in several recent papers
(for example \cite{Des94} and references therein). The daughter nucleus
$^{17}$C, in turn, is also a multi-neutron emitter \cite{Ree91}, but with
lower probability. Current literature data for $^{17}$C show
two low-lying levels, with excitation energies of 292 keV and 295 keV,
respectively. Three closely spaced low-lying states are expected \cite{Til93}.
Until now, however, no data on higher lying levels in $^{17}$C, in particular,
any levels close to the
neutron threshold at 730 keV \cite{Til93}, have been available.
With the present experiment we have attempted to increase appreciably the
information available about $^{17}$C levels, especially those just above the
$^{16}$C + n threshold. These would enable the (n,$\gamma$)-reaction to
proceed resonantly.
Thus the results are of interest from a purely nuclear
physics perspective, but also for that of astrophysics and the study of IMs of
primordial nucleosynthesis.

\section{Experimental setup}

The experiment was done at the projectile fragmentation type
radioactive beam facility (RIPS) of the Institute
of Physical and Chemical Research (RIKEN). A description of the facility
is given in \cite{Kub92}. The setup was identical to that used to study
the $\beta$-delayed neutron decay of $^{19}$C\cite{Oza95}.
Briefly, a primary beam of $^{22}$Ne with an energy
of 110 MeV/nucleon impinged on a primary target of $^9$Be (thickness 5 mm).
An ECR ion source produced the initial beam, and the acceleration
of the ions was performed by the AVF Cyclotron and the RIKEN Ring Cyclotron.
The primary beam intensity at the target was about 55 pnA, which
resulted in a $^{17}$B rate of about 120 s$^{-1}$.
A schematic diagram of the setup is given in Fig.\ \ref{fig1}.
The projectile fragments were analysed by two dipole magnets,
which were adjusted to
maximize the rate of $^{17}$B ions. An Al-wedge was included at
the primary focus F1 to disperse beam contaminants. Particle identification
during beam tuning and focusing was done using time-of-flight
and $\Delta$E detection. Adjustable slits at the focal points F1, F2
and F3 were used to increase the purity of the beam. The $^{17}$B ions
passed through a thin mylar window (25 $\mu$m) and through a final rotatable
Al-absorber (675 mg/cm$^2$, thickness 2.5 mm) which degraded the energy
of the ions sufficiently for them to be stopped in a plastic scintillator.
The $^{17}$B ions then underwent $\beta$-decay into
$^{17}$C (half-life of about 5 ms \cite{Til93}, $Q_\beta$-value of
22.68(14) MeV \cite{Wap93}).
The $\beta$-decay was followed by a one-neutron decay
with a probability of $p_{1n} = 63(1)$\% \cite{Duf88}
(the probability
for decay into the $^{17}$C ground state without emission of a neutron,
$p_{0n}$, is $21(2)$\%, and $p_{2n} = 11(7)$\%). The energies of
the emitted neutrons were determined by time-of-flight (TOF), with the
$\beta$-decay providing the start signal, and large neutron walls yielding the
stop signal. In addition, NaI counters were used to detect
$\gamma$-events from decays
of excited $^{16}$C states in cases in which the neutron decay was not to the
$^{16}$C ground state.
The beam purity was high, the main contaminant being $^{19}$C
(the ratio $^{17}$B/$^{19}$C was about 10/1). No neutron groups stronger
than the background were observed from $^{19}$C $\beta$-delayed
neutron decay. During the experiment the
yield at F3 was typically 20 to 40 s$^{-1}$.

The $\beta$-detector consisted of an assembly of 6 plastic
scintillator components, labeled S$_1$, S$_2$, A, S$_3$, S$_4$ and E,
from left to right in Fig.\ \ref{fig1}.
The detector arrangement was tilted 45$^\circ$ with respect to the beam axis in
order to minimize the length of neutron passage through the scintillator.
The $^{17}$B ions were implanted and stopped in
A. The S-counters had dimensions of 10 cm $\times$ 10 cm, with a thickness of
2 mm. Counter A was 4 cm $\times$ 6 cm, with 7.2 mm thickness. The E-counter
was a block of 10 cm $\times$ 10 cm $\times$ 10 cm and was used to observe the
$\beta$-spectra.

Detection of the neutrons was achieved by three large neutron walls, each
having an active surface area of about 100 cm $\times$ 100 cm, the centers of
which were positioned at a distance of 125 cm from detector A. This translates
to a total solid angle of about 1.4 sr. Each wall consisted of 15 or 16 plastic
scintillator bars (scintillator type BC-408), with dimensions 6 cm $\times$ 6
cm $\times$ 100 cm, with a Hamamatsu photomultiplier (model R329-02) attached
to each end. The time resolution of the neutron walls was about 1.5 ns, and the
energy resolution for 1 MeV neutrons was about 4.6\% (both resolutions in
$\sigma$ of a Gaussian).

For detection of $\gamma$-events a 7.6 cm $\Phi \times$ 7.6 cm NaI(Tl)
scintillator was mounted at a distance of 16 cm from counter A (corresponding
to a solid angle of 1.0 sr).

The beam was operated in a pulsed mode, with beam-on for 10 ms (two half-lives
of $^{17}$B), and beam-off (measuring time window) for 20 ms, with data
acquisition occuring only in the beam-off mode. After decay
of a $^{17}$B in the A-counter, the energy of each $\beta$-delayed neutron
was measured by TOF between the A-component of the $\beta$-counter (start) and
the neutron walls (stop). A triple $\beta$-coincidence
either between counters A, S$_1$ and S$_2$, or between counters
A, S$_3$ and S$_4$ was required for a valid start signal.

All the photomultipliers in the neutron walls were adjusted roughly to the
same gain. The gain was monitored periodically by bringing
$\gamma$-sources close to each wall ($^{60}$Co and $^{137}$Cs) and observing
the Compton edges of the pulse-height spectra. The NaI-counter was calibrated
with $^{60}$Co, $^{137}$Cs and $^{22}$Na sources.

\section{Data reduction and results}

The neutron time-of-flight spectra were derived in the standard way.
For each scintillator bar, the TOF was determined by averaging the
timing signals from the two photomultipliers.
The signals were then corrected as a function of their amplitudes,
and for different path lengths of neutrons hitting different regions of the
walls, as determined from the difference of the timing signals from the
two photomultipliers.

The neutron TOF values were measured with an
internal clock. In order to get an accurate time calibration,
the beam was switched to $^{17}$N, and neutrons from the $\beta$-delayed
decay of $^{17}$N into $^{16}$O + n were observed (Fig.\ \ref{fig1a}).
Three peaks with the known
neutron lab-energies (branchings) \cite{Til93} of 1.70 MeV (6.9\%),
1.17 MeV (50.1\%) and 0.38 MeV (38.0\%) were observed. These neutron energies
correspond to level energies above threshold \cite{Til93}
of 1.81 MeV, 1.24 MeV and 0.41 MeV.
A fourth known peak (neutron energy 0.88 MeV, energy above threshold 0.94 MeV,
branching 0.6\%) was much too weak and too close to the 1.17 MeV peak to be
observed. The three peaks were used for the neutron energy
calibration. Since the 0.38 MeV peak was quite strong, this indicates that the
neutron detection threshold was well below that energy.

Fig.\ \ref{fig1a} shows that the $^{17}$N TOF peaks are clearly asymmetric.
Although we do not fully understand the cause of the asymmetry, we assumed that
all events under the tail of each asymmetric peak belonged to that peak.
Thus an asymmetric fit function was
used to obtain each integral. The asymmetric function (a combination of a
Gaussian plus a tail for the larger TOF side consisting of a polynomial
multiplied by an exponential term) was obtained
from the prominent, and isolated,  1.17 MeV peak. In subsequent fits, the
same asymmetric
shape was consistently applied to all peaks (the tail was matched to the
Gaussian at the half maximum point, and its width was scaled with the
width of the Gaussian); the only free parameters then
were the three parameters of the underlying
Gaussian. The background was fitted
with a broad Gaussian. It should be pointed out that the exact (asymmetric)
shape is not crucial for obtaining the integrals, as long as all peaks are
fitted {\em consistently} using the same shape. Indeed, using Gaussians
or Lorentzians for fitting gives the same relative intensities within error
bars.

In the data analysis, only 31 of the 47 bars in the three neutron walls were
used. These were the ones with the lowest noise phototubes, hence, those that
allowed the lowest threshold settings. The thresholds of these detectors were
verified to be roughly 15 keV electron-equivalent energy. Because this setting
is so low, the efficiencies of all the detectors used in the analysis were
essentially the same for the neutron energies we observed, as was verified by
comparing neutron yields, especially those at low energies, observed in higher
and lower threshold groups of detectors. The absolute efficiency of the
detectors used (see Fig.\ \ref{fig2}) was calculated by means of a Monte-Carlo
simulation using a standard prescription \cite{Cec79}. The accuracy of the
calculated efficiencies was then checked by comparing the yields in the peaks
from the $^{17}$N $\beta$-delayed neutron emission with the well established
branching ratios of that decay. As can be seen from Table \ref{tab0}, the
branching ratios determined from our data, using our calculated efficiencies,
are in excellent agreement with the known ratios, thus confirming the accuracy
of our efficiencies to within a few percent down to a neutron energy of 0.38
MeV. Figure \ref{fig3} shows a fully corrected $^{17}$B $\beta$-delayed neutron
TOF spectrum. The peaks in this figure were fitted simultaneously with the
asymmetric shape described above. A broad Gaussian was used for the background
in order to account for two-neutron breakup.

The three main peaks correspond to
neutron energies of 2.91(5) MeV, 1.80(2) MeV and 0.82(1) MeV (Table
\ref{tab1}).
An investigation of
possible daughter decays (see Figure \ref{fig4} for a level diagram)
reveals that the 0.82 MeV peak corresponds to
the $\beta$-delayed neutron decay of $^{16}$C into $^{15}$N + n.
According to published data \cite{Til93}, there are two $^{16}$N states
above threshold that decay into the neutron channel (level energies in
$^{16}$N of 3.36 MeV and 4.32 MeV). These states give rise to neutron
energies (after recoil correction) of 0.81 MeV (branching ratio of 84\%) and
1.72 MeV (branching ratio of 16\%). The 1.72 MeV neutron group could not be
seen
separately in the spectrum because it was under the shoulder of the much
stronger peak at 1.80 MeV. The identification of the 0.81 MeV peak is
confirmed by the correlation of its intensity in the TOF spectrum with
gates of different
$\beta$-energies. The $\beta$s from $^{17}$B leading to
the two observed neutron peaks have maximum energies of about 20 MeV.
The $\beta$s from the decay of $^{16}$C leading to the third observed
neutron peak, on the other hand, have maximum energies of only about 5 MeV
(Fig.\ \ref{fig4}). Hence, gating the neutron TOF spectrum with only the high
energy portion of the $\beta$-spectrum should make the $\beta$-delayed
$^{16}$C peak disappear, as was observed.
The remaining two main neutron peaks are consistent with decays of excited
$^{17}$C states, both from the time scale of associated $\beta$-decays and
from the maximum $\beta$-energies.

The broad shoulder under the 1.81 MeV peak requires special attention, since
the asymmetry of this peak is greater than that of the (isolated) 1.17 MeV peak
in $^{17}$N. This is a clear indication that the broad shoulder is actually
caused by one or more smaller unresolved peaks. From the known branching ratios
of the $^{16}$C peaks at 0.81 MeV and 1.72 MeV, the known relative intensity of
the 0.81 MeV peak in the TOF spectrum and the known efficiency of the
detectors, the 1.72 MeV peak (assuming it has a comparable Gaussian width) is
fully determined and can be included in the fit. This new fit still does not
fully account for the asymmetry of the 1.81 MeV peak. The remaining shoulder of
the 1.81 MeV peak is apparently due to another neutron group at 1.43 MeV from
$\beta$-delayed $^{17}$B decay (Table \ref{tab1}).

Another possible neutron peak, at 0.42 MeV, can be observed only with
difficulty in the TOF spectrum, since it is very broad there and cannot easily
be distinguished from the background. However, it becomes more apparent when
the TOF spectrum is transformed into an energy spectrum (Fig.\ \ref{fig7}).
Note that the peak seen in Fig.\ \ref{fig7} cannot be a manifestation of a
background subtraction; the energy spectrum in Fig.\ \ref{fig7} is the
transformation of the raw TOF spectrum. The centroid for this peak was first
determined from a fit on the energy scale and was then included in the fit of
the TOF spectrum. Due to the poor peak-to-background ratio even in the energy
spectrum, not much can be said about this peak. Since it is so broad, it may
well be a doublet. Its width might also imply it results from a more highly
excited state in $^{17}$C that decays to an excited state in $^{16}$C. However,
the data obtained for this state do not allow such a determination.

Both the 1.43 MeV peak and the 0.42 MeV peak are consistent with decays
of excited $^{17}$C states, both from the time scale of associated
$\beta$-decays and from the maximum $\beta$-energies.

Other neutrons could possibly come from decays of the daughters $^{15}$C and
$^{17}$C. The $^{15}$C ground state is above the neutron
threshold of $^{15}$N. In the $^{17}$C case, the $\beta$-delayed neutron
emission probability $p_n$ has been reported\cite{Ree91,Ree94},
and very recently also the $\beta$-branchings to excited states in $^{17}$N
\cite{Sch95}. The subsequent $\beta$-decay and
neutron decays of $^{17}$N, in turn, were studied in the energy calibration
run with $^{17}$N. No indication of $^{17}$C or $^{17}$N peaks is apparent in
the $^{17}$B neutron TOF spectrum.

The $\gamma$-spectrum is weak and exhibits no clear peaks.
Coincidences between $\gamma$-rays and each of the neutron peaks resulting
from neutron decay of $^{17}$C also do not produce peaks.
Unfortunately, the gain for the $\gamma$-spectrum was set so that the
1.77 MeV $\gamma$-rays from the $^{16}$C first excited state
would have ocurred at the high energy edge of the $\gamma$-spectrum.
However, that spectrum was flat near its high energy end, suggesting
that all $^{17}$C neutron decays were to the $^{16}$C ground state.
Fortunately, Kurie plots can also be used to determine the $^{17}$C
states to which the $^{17}$B $\beta$-decays, and hence whether the
energies of the resulting neutrons are consistent with decays to the
$^{16}$C ground state or first excited state. This technique, however,
works only for strong well-separated peaks in the neutron TOF spectrum,
since the neutron peaks are used as gates for the $\beta$-spectrum.
Since the neutron and $\beta$-spectrum end point energies from $^{17}$N and
$^{16}$C decays are known and those from the $\beta$-delayed $^{19}$C decays
could be determined unambiguously (see \cite{Oza95}), the calibration
between channels and maximum $\beta$-energies can be used to calibrate the
$\beta$-spectra from $^{17}$B decays to $^{17}$C states. If it is assumed
that the two well-separated neutron peaks result from decays to the $^{16}$C
ground state, a good fit to the seven data points is achieved, whereas
assumption of decay to the first excited state, or any other excited state,
produces a much worse fit. This, together with the fact that the
$\gamma$-spectrum is flat near the threshold at 1.77 MeV indicates that all
neutron decays of $^{17}$C populate only the $^{16}$C ground state.

Table \ref{tab1} lists the assigned energies of all neutron peaks, as
well as the branching ratios and log($ft$) values. The branching ratios
were obtained by normalizing to the $p_{1n}$-value given in \cite{Duf88}.
More recent experimenters \cite{Ree91,Lew89} have
measured only the $\beta$-delayed neutron emission probability $p_n$
summed over one- and multi-neutron decay. These literature values all agree
within their error bars. A new investigation \cite{Ree94}, however, results
in a somewhat higher value for $p_n$.

Figure \ref{fig6} shows one of the measured $\beta$-time spectra for $^{17}$B,
gated with the 1.80 MeV neutron group from the decay of $^{17}$C into
$^{16}$C + n. It was fitted with one exponential component.
The $\beta$-time spectrum gated with $\beta$-delayed neutrons from the
decay of $^{16}$C could not be fitted to give the half-life of
$^{16}$C, since this half-life, 747 ms \cite{Til93}, is so much
longer than our time window (20 ms).
The calculated half-life for $^{17}$B, averaged over several
different fits, is 4.93(19) ms, in reasonable agreement with
a weighted average half-life of 5.10(5) ms based on literature values
(5.3(6) ms, 5.08(5) ms and 5.9(3) ms given in \cite{Til93}, and
5.20(45) ms given in \cite{Ree94}).

An attempt was made to identify resonant two-neutron decay. To do so, we first
rejected events in which the second neutron signal out of a pair was detected
in an adjacent neutron bar. In such cases, the second neutron event was most
likely initiated by the same (scattered) neutron as the first event. Since the
remaining two-neutron events had very poor statistics, the results were not
conclusive.

{}From systematic studies of $\beta$-decay, all the observed log($ft$) values
appear to be consistent with allowed transitions (although first forbidden
transitions cannot be totally ruled out). With a $^{17}$B
ground-state spin/parity of $J^\pi = 3/2^-$\cite{Til93}, the spin/parities
of the observed $^{17}$C states, assuming allowed transitions, are
restricted to $(1/2,3/2,5/2)^-$. Due to the selection
rules for allowed $\beta$-decay ($\Delta J = 0, \pm 1$, no change of parity)
only negative parity states were observed. Positive parity states could
be populated by (first order) forbidden transitions,
which would generally be expected to have lifetimes at least an order
of magnitude longer. With the present setup (assuming the correctness
of the assumption of allowed transitions), we would not have detected such
forbidden transitions, and we therefore cannot exclude the presence of
possible low-lying positive parity states in $^{17}$C.
Theoretical log($ft$) values and spin assignments based on the shell model
are in preparation \cite{Kit95}.

Within the limits of resolution and detection threshold, this experiment
provides no evidence for negative parity states close to the neutron
threshold. The observed states in $^{17}$C have too high excitation energies
to be of astrophysical relevance in IM scenarios. Any
$^{16}$C(n,$\gamma$)$^{17}$C capture reaction at low energies would therefore
have to proceed as a direct capture, as has been assumed in previous IM
calculations. Experiments sensitive to positive parity states would clearly
be required to provide a final answer.

\acknowledgements
We wish to thank the staff of the RIKEN cyclotron for the smooth and
reliable operation of all equipment. The American contingent of the authors
is grateful for the warm hospitality at RIKEN. GR acknowledges support through
an OSU postdoctoral fellowship. This work was supported by NSF grants
PHY-9221669, INT-9218241 and PHY-9100688.


%
\newpage


\begin{figure}
\caption{Schematic diagram of the experimental setup. The
$^{17}$B beam enters from the left, passes through the rotatable energy
degrader, the first two scintillators S$_1$ and S$_2$ and is stopped in the
implantation $\beta$-detector A. Also shown in the diagram are the E-detector
used for measuring the $\beta$-spectra, the NaI $\gamma$-ray detector and the
neutron walls. }
\label{fig1}
\end{figure}

\begin{figure}
\caption{Fully corrected $^{17}$N $\beta$-delayed neutron TOF spectrum
used for the energy calibration.
The channel numbers correspond to the time-of-flight in ns.
}
\label{fig1a}
\end{figure}

\begin{figure}
\caption{Neutron efficiency for the detectors used in the data analysis
as a function of the neutron energy, calculated for neutron energies
above 0.3 MeV. }
\label{fig2}
\end{figure}

\begin{figure}
\caption{Fully corrected $^{17}$B $\beta$-delayed neutron TOF spectrum.
The channel numbers correspond to the time-of-flight in ns.
The $^{17}$B
peaks are observed at neutron lab-energies of 2.91(5) MeV, 1.80(2) MeV,
1.43(2) MeV and 0.42(1) MeV. The 1.72 MeV and 0.82 MeV peaks
correspond to neutron decays in the $\beta$-delayed decay of the daughter
nucleus $^{16}$C. }
\label{fig3}
\end{figure}

\begin{figure}
\caption{Level and decay scheme of $^{17}$B
\protect\cite{Duf88,Til93,Wap93,Ajz86}. The dashed lines are the new levels. }
\label{fig4}
\end{figure}

\begin{figure}
\caption{$^{17}$B $\beta$-delayed neutron TOF spectrum, transformed to an
energy scale. Each channel corresponds to 20 keV.
The peak at 0.42 MeV is clearly visible. }
\label{fig7}
\end{figure}

\begin{figure}
\caption{Time spectrum of $^{17}$B $\beta$-decay, gated to the 1.80 MeV neutron
peak. The channel numbers correspond to the time in 0.01 ms.
The solid line is an exponential fit. }
\label{fig6}
\end{figure}


\begin{table}[htbp]
\caption{Branching ratios obtained from the $\beta$-delayed
neutron decay of $^{17}$N, compared to literature values \protect\cite{Til93}.
The uncertainties of the literature energies are in the 1-keV range. The main
contribution to the errors in the branching ratios results from the uncertainty
of the $\beta$-efficiencies caused by uncertainties of the threshold. }
\label{tab0}
\begin{center}
\begin{tabular}{ccc}
$E_n$ (MeV) & branching (\%) 	& lit.\ branching (\%)	\\ \tableline
1.70	&	6.4(10)	 \tablenotemark[1]	&	6.9(5)	 	\\
1.17	&	49.1(46) \tablenotemark[1]	&	50.1(13) 	\\
0.38	&	39.5(46) \tablenotemark[1]	&	38.0(13) 	\\
\end{tabular}
\vspace{-0.5cm}
\tablenotetext[1]{Normalized to 95(1) \% \cite{Til93} }
\end{center}
\end{table}

\begin{table}[htbp]
\caption{Energy assignments, branching ratios (normalized to
the known probability for 1n-decay) and log $ft$-values for observed
$\beta$-delayed neutrons from $^{17}$B. }
\label{tab1}
\begin{center}
\begin{tabular}{cccccc}
$E_n$ (MeV) & $E_x$ (MeV) & in & decay to &
				  branching (\%) & $\log{ft}$ \\
\tableline
2.91(5) & 3.82(5) & $^{17}$C & g.s. & 17.8(14) \tablenotemark[1] & 4.91(15)  \\
1.80(2) & 2.64(2) & $^{17}$C & g.s. & 29.6(25) \tablenotemark[1] & 4.82(14)  \\
1.72    & 4.32    & $^{16}$N \tablenotemark[2] & g.s. \\
\mbox{[} 1.43(2) \tablenotemark[4] & 2.25(2) & $^{17}$C & g.s. &
                     4.7(15) \tablenotemark[1] & 5.66(19) \mbox{]} \\
0.82(1) & 3.37(1) \tablenotemark[3] & $^{16}$N & g.s. \\
\mbox{[} 0.42(1) \tablenotemark[5] & 1.18(1) & $^{17}$C & g.s. &
                    10.8(30) \tablenotemark[1] & 5.41(18) \mbox{]} \\
\end{tabular}
\vspace{-0.5cm}
\tablenotetext[1]{Normalized to 63(1) \% \cite{Duf88} }
\tablenotetext[2]{known peak with known branching \cite{Til93} included in fit}
\tablenotetext[3]{$E_x$ = 3.36 \cite{Til93} }
\tablenotetext[4]{peak is under broad shoulder; assignment suggested by peak
fitting}
\tablenotetext[5]{suggested by energy spectrum}
\end{center}
\end{table}


\end{document}